\documentclass[12pt]{article}
\newcommand{\mc}{m_{\rm c}}
\newcommand{\gev}{\, {\rm GeV}}
\newcommand{\mev}{\, {\rm MeV}}
\newcommand{\Lms}{\Lambda_{\overline{\rm MS}}}

\newcommand{\as}{\alpha_s}
\newcommand{\ord}{{\cal O}}
 
\begin{document}
 
\thispagestyle{empty}
\begin{flushright}
CERN-TH/98-369\\
TUM-T31-337/98\\
December 1998 
\end{flushright}

\vspace*{1.5cm}
\centerline{\Large\bf The Rare Decays $K\to\pi\nu\bar\nu$,
              $B\to X\nu\bar\nu$ and $B\to l^+l^-$:}
\vspace*{0.3cm}
\centerline{\Large\bf An Update}
\vspace*{2cm}
\centerline{{\sc Gerhard Buchalla${}^a$} and {\sc Andrzej J. Buras${}^b$}}
\bigskip
\centerline{\sl ${}^a$Theory Division, CERN, CH-1211 Geneva 23,
                Switzerland}
\vspace*{0.2cm}
\centerline{\sl ${}^b$Physik Department, Technische Universit{\"a}t
                   M{\"u}nchen,}
\centerline{\sl D-85748 Garching, Germany}
 
\vspace*{1.5cm}
\centerline{\bf Abstract}
\vspace*{0.3cm}

We update the Standard Model predictions for the rare decays
$K^+\to\pi^+\nu\bar\nu$ and $K_L\to\pi^0\nu\bar\nu$. In view
of improved limits on $B_s$--$\bar B_s$ mixing we derive a
stringent and theoretically clean Standard Model upper limit
on $B(K^+\to\pi^+\nu\bar\nu)$, which is based on
the ratio of $B_d$--$\bar B_d$ to $B_s$--$\bar B_s$ mixing,
$\Delta M_d/\Delta M_s$, alone. This method avoids the large
hadronic uncertainties present in the usual analysis of the 
CKM matrix. We find $B(K^+\to\pi^+\nu\bar\nu)< 1.67\cdot 10^{-10}$,
which can be further improved in the future.
In addition we consider the extraction of $|V_{td}|$ from a 
future measurement of $B(K^+\to\pi^+\nu\bar\nu)$, discussing the 
various sources of uncertainties involved.
We also investigate theoretically clean constraints on
$B(K_L\to\pi^0\nu\bar\nu)$.
We take the opportunity to review the next-to-leading
order (NLO) QCD corrections to $K\to\pi\nu\bar\nu$, $K_L\to\mu^+\mu^-$,
$B\to X\nu\bar\nu$ and $B\to l^+l^-$, including a small additional term 
that had been missed in the original publications. The phenomenological
impact of this change is negligible, the corresponding numerical
shift being essentially within the small perturbative uncertainties
at the NLO level.

\noindent 
\vfill

 
\newpage
\pagenumbering{arabic}

\section{Introduction}

The rare decays $K^+\to\pi^+\nu\bar\nu$, $K_L\to\pi^0\nu\bar\nu$, 
$B\to X_{s,d}\nu\bar\nu$ and $B_{s,d}\to l^+l^-$ are very promising
probes of flavour physics. They are sensitive to the
quantum structure of Standard Model flavour dynamics and can
at the same time be computed theoretically to an exceptionally
high degree of precision.
In particular the kaon modes
$K^+\to\pi^+\nu\bar\nu$ and $K_L\to\pi^0\nu\bar\nu$
have received considerable interest in recent years.
The theoretical status of these decays has been reinforced by
the calculation of next-to-leading order QCD corrections
\cite{BB1}--\cite{BB6}.
The corrections to the top-quark dominated modes
$K_L\to\pi^0\nu\bar\nu$, $B\to X_{s,d}\nu\bar\nu$ and 
$B_{s,d}\to l^+l^-$ \cite{BB1,BB2} have recently been recomputed 
in \cite{MU}. The results of \cite{MU} confirm the calculations
of \cite{BB1,BB2} up to a small term arising from the box
diagram that had been overlooked
in \cite{BB2}. We are in agreement with \cite{MU} that this
term needs to be included to obtain the complete NLO correction.
As will be explained in more detail below, the missing piece
follows from a simple one-loop calculation, which can be 
performed separately, and it can be added to the old
result. The latter was obtained from the two-loop matching calculation
in \cite{BB2}, which remains unaffected.

In contrast to $K_L\to\pi^0\nu\bar\nu$, where only the top-quark
contribution is relevant, also the charm sector is important
for $K^+\to\pi^+\nu\bar\nu$. This sector, which has not been 
considered in \cite{MU}, will also be discussed in the present
article.

The numerical impact of the modification is small.
The changes are within the perturbative error of the NLO calculation,
both for the top and the charm contribution. 
All qualitative conclusions, regarding the NLO error analysis
from residual scale dependence and the theoretical precision
that can be achieved in the calculation of these decays, remain unchanged. 

The purpose of this paper is twofold. We will first present the
complete NLO effective hamiltonians for $K^+\to\pi^+\nu\bar\nu$
(section 2), the short-distance part of $K_L\to\mu^+\mu^-$
(section 3) and the top-quark dominated modes
$K_L\to\pi^0\nu\bar\nu$, $B\to X_{s,d}\nu\bar\nu$ and
$B_{s,d}\to l^+l^-$ (section 4), including the additional
NLO contribution from the box diagram. The origin and the derivation
of this contribution are explained in the Appendix.
Secondly, we update the Standard Model predictions for
$B(K^+\to\pi^+\nu\bar\nu)$ and $B(K_L\to\pi^0\nu\bar\nu)$
and discuss some phenomenological aspects
of these decays that have not been emphasized before.
In particular, we derive in section 5 a clean correlation between the 
ratio of $B_d$--$\bar B_d$ and $B_s$--$\bar B_s$ mixing and the
$K^+\to\pi^+\nu\bar\nu$ branching fraction. In view of recent
experimental progress, both for $K^+\to\pi^+\nu\bar\nu$ and $B$--$\bar B$
mixing, this could provide an interesting Standard Model test in
the near future. We also analyse the determination of $|V_{td}|$
from $B(K^+\to\pi^+\nu\bar\nu)$ and investigate the theoretical
uncertainties involved. 
Section 6 explores the impact of $\Delta M_d/\Delta M_s$ and
$\sin 2\beta$, as obtained from the CP asymmetry in
$B_d(\bar B_d)\to J/\Psi K_S$, on the neutral mode 
$K_L\to\pi^0\nu\bar\nu$.
The combination of these three theoretically clean observables
offers another stringent consistency check on the Standard Model.
We conclude in section 7.

\section{Effective Hamiltonian for $K^+\to\pi^+\nu\bar\nu$}

The effective hamiltonian for $K^+\to\pi^+\nu\bar\nu$ can be written as
\begin{equation}\label{hkpn} {\cal H}_{eff}={G_F \over{\sqrt 2}}{\alpha
\over 2\pi \sin^2\Theta_W}
 \sum_{l=e,\mu,\tau}\left( V^{\ast}_{cs}V_{cd} X^l_{NL}+
V^{\ast}_{ts}V_{td} X(x_t)\right)
 (\bar sd)_{V-A}(\bar\nu_l\nu_l)_{V-A} \, .
\end{equation}
The index $l$=$e$, $\mu$, $\tau$ denotes the lepton flavour.
The dependence on the charged lepton mass, resulting from the box-graph,
is negligible for the top contribution. In the charm sector this is the
case only for the electron and the muon, but not for the $\tau$-lepton.
\\
The function $X(x)$, relevant for the top part, reads
to $\ord(\as)$ and to all orders in $x=m^2/M^2_W$
\begin{equation}\label{xx} 
X(x)=X_0(x)+\frac{\as}{4\pi} X_1(x) 
\end{equation}
with \cite{IL}
\begin{equation}\label{xx0} X_0(x)={x\over 8}\left[ -{2+x\over 1-x}+
{3x-6\over (1-x)^2}\ln x\right] \end{equation}
and the QCD correction 
\begin{eqnarray}\label{xx1}
X_1(x)=&-&{29x-x^2-4x^3\over 3(1-x)^2}-{x+9x^2-x^3-x^4\over (1-x)^3}\ln x
\nonumber\\
&+&{8x+4x^2+x^3-x^4\over 2(1-x)^3}\ln^2 x-{4x-x^3\over (1-x)^2}L_2(1-x)
\nonumber\\
&+&8x{\partial X_0(x)\over\partial x}\ln x_\mu
\end{eqnarray}
where $x_\mu=\mu^2/M^2_W$ with $\mu=\ord(m_t)$ and
\begin{equation}\label{l2} L_2(1-x)=\int^x_1 dt {\ln t\over 1-t}   
\end{equation}
The $\mu$-dependence in the last term in (\ref{xx1}) cancels to the
order considered the $\mu$-dependence of the leading term $X_0(x(\mu))$.
\\
The expression corresponding to $X(x_t)$ in the charm sector is the function
$X^l_{NL}$. It results from the RG calculation in NLLA and is given
as follows:
\begin{equation}\label{xlnl}
X^l_{NL}=C_{NL}-4 B^{(1/2)}_{NL}  
\end{equation}
$C_{NL}$ and $B^{(1/2)}_{NL}$ correspond to the $Z^0$-penguin and the
box-type contribution, respectively. One has 
\begin{eqnarray}\label{cnln}
\lefteqn{C_{NL}={x(m)\over 32}K^{{24\over 25}}_c\left[\left({48\over 7}K_++
 {24\over 11}K_--{696\over 77}K_{33}\right)\left({4\pi\over\as(\mu)}+
 {15212\over 1875} (1-K^{-1}_c)\right)\right.}\nonumber\\
&&+\left(1-\ln{\mu^2\over m^2}\right)(16K_+-8K_-)-{1176244\over 13125}K_+-
 {2302\over 6875}K_-+{3529184\over 48125}K_{33} \nonumber\\
&&+\left. K\left({56248\over 4375}K_+-{81448\over 6875}K_-+{4563698\over 
144375}K_{33}
  \right)\right]
\end{eqnarray}
where
\begin{equation}\label{kkc} K={\as(M_W)\over\as(\mu)}\qquad
  K_c={\as(\mu)\over\as(m)}  \end{equation}
\begin{equation}\label{kkn} K_+=K^{{6\over 25}}\qquad K_-=K^{{-12\over 25}}
\qquad      K_{33}=K^{{-1\over 25}}  \end{equation}
\begin{eqnarray}\label{bnln}
\lefteqn{B^{(1/2)}_{NL}={x(m)\over 4}K^{24\over 25}_c\left[ 3(1-K_2)\left(
 {4\pi\over\as(\mu)}+{15212\over 1875}(1-K^{-1}_c)\right)\right.}\nonumber\\
&&-\left.\ln{\mu^2\over m^2}-
  {r\ln r\over 1-r}-{77\over 3}+{15212\over 625}K_2+{4364\over 1875}K K_2
  \right]
\end{eqnarray}
Here $K_2=K^{-1/25}$, $m=m_c$, $r=m^2_l/m^2_c(\mu)$ and $m_l$ is the
lepton mass.  We will at times omit the index $l$ of $X^l_{NL}$.  In
(\ref{cnln}) -- (\ref{bnln}) the scale is $\mu=\ord(m_c)$.  For the
charm contribution we need the
two-loop expression for $\as(\mu)$ given by  
\begin{equation}\label{asmu}
\alpha_s(\mu)=\frac{4\pi}{\beta_0\ln\frac{\mu^2}{\Lambda^2}}
\left[ 1-\frac{\beta_1}{\beta^2_0}
\frac{\ln \ln\frac{\mu^2}{\Lambda^2}}{\ln\frac{\mu^2}{\Lambda^2}}
\right]
\end{equation}
\begin{equation}\label{b01}
\beta_0=11-\frac{2}{3}f\qquad\quad \beta_1=102-\frac{38}{3}f
\end{equation}
The effective number of flavours is $f=4$ in the expressions
above for the charm sector. The QCD scale in (\ref{asmu}) is
$\Lambda=\Lambda^{(4)}_{\overline{MS}}=(325\pm 80)\,{\rm MeV}$.

Again -- to
the considered order -- the explicit $\ln(\mu^2/m^2)$ terms in
(\ref{cnln}) and (\ref{bnln}) cancel the $\mu$-dependence of the
leading terms.

For phenomenological applications it is useful to define
for the top contribution a QCD correction factor $\eta_X$ by
\begin{equation}\label{xxt}
X(x_t)=\eta_X\cdot X_0(x_t) \qquad \eta_X=0.994
\end{equation} 
With the ${\overline{MS}}$ definition of the top-quark mass,
$m_t\equiv\bar m_t(m_t)$, the QCD factor $\eta_X$ is practically
independent of $m_t$. This is the definition of $m_t$ that we
will employ throughout this paper.

For the charm sector it is useful to introduce
\begin{equation}\label{p0k}
P_0(X)=\frac{1}{\lambda^4}\left[\frac{2}{3}X^e_{NL}+
 \frac{1}{3}X^\tau_{NL}\right]
\end{equation}
with the Wolfenstein parameter \cite{WO} $\lambda=0.22$.
Results for the charm functions are collected in tables 
\ref{tab:xnlnum} and \ref{tab:P0Kplus},
where $m_c$ stands for the ${\overline{MS}}$ mass $\bar m_c(m_c)$.

\begin{table}[htb]
\caption[]{The functions $X^e_{\rm NL}$ and $X^\tau_{\rm NL}$
for various $\Lms^{(4)}$ and $\mc$.
\label{tab:xnlnum}}
\begin{center}
\begin{tabular}{|c|c|c|c|c|c|c|}
\hline
& \multicolumn{3}{c|}{$X^e_{\rm NL}/10^{-4}$} &
  \multicolumn{3}{c|}{$X^\tau_{\rm NL}/10^{-4}$} \\
\hline
$\Lms^{(4)}\ [\mev]\;\backslash\;\mc\ [\gev]$ &
1.25 & 1.30 & 1.35 & 1.25 & 1.30 & 1.35 \\
\hline
245 & 10.73  & 11.60  & 12.50 & 7.34 & 8.06 & 8.81 \\
285 & 10.44  & 11.31  & 12.20 & 7.05 & 7.77 & 8.52 \\
325 & 10.14  & 11.00  & 11.90 & 6.75 & 7.47 & 8.21 \\
365 &  9.82  & 10.69  & 11.58 & 6.44 & 7.15 & 7.89 \\
405 &  9.49  & 10.35  & 11.24 & 6.10 & 6.81 & 7.55 \\
\hline
\end{tabular}
\end{center}
\end{table}

\begin{table}[htb]
\caption[]{The function $P_0(X)$ for various $\Lms^{(4)}$ and $m_c$.
\label{tab:P0Kplus}}
\begin{center}
\begin{tabular}{|c|c|c|c|}
\hline
&\multicolumn{3}{c|}{$P_0(X)$}\\
\hline
$\Lms^{(4)}\ [\mev]\;\backslash\;\mc\ [\gev]$ & 
$1.25$ & $1.30$ & $1.35$  \\
\hline
$245$ & 0.410 & 0.445 & 0.481 \\
$285$ & 0.397 & 0.432 & 0.468 \\
$325$ & 0.385 & 0.419 & 0.455 \\
$365$ & 0.371 & 0.406 & 0.442 \\
$405$ & 0.357 & 0.391 & 0.427 \\ 
\hline
\end{tabular}
\end{center}
\end{table}

\section{Effective Hamiltonian for $(K_L\to\mu^+\mu^-)_{SD}$}

The analysis of $(K_L\to\mu^+\mu^-)_{SD}$ proceeds in essentially the same
manner as for $K^+\to\pi^+\nu\bar\nu$. 
The only difference is introduced through the
reversed lepton line in the box contribution. In particular there is
no lepton mass dependence, since only massless neutrinos appear as
virtual leptons in the box diagram.\\
The effective hamiltonian in next-to-leading order can be written as
follows:
\begin{equation}\label{hklm}{\cal H}_{eff}=
-{G_F \over{\sqrt 2}}{\alpha\over 2\pi \sin^2\Theta_W}
 \left( V^{\ast}_{cs}V_{cd} Y_{NL}+
V^{\ast}_{ts}V_{td} Y(x_t)\right)
 (\bar sd)_{V-A}(\bar\mu\mu)_{V-A} + h.c. \end{equation}
The function $Y(x)$ is given by
\begin{equation}\label{yy}
Y(x) = Y_0(x) + \frac{\alpha_s}{4\pi} Y_1(x)\end{equation}
where \cite{IL}
\begin{equation}\label{yy0}
Y_0(x) = {x\over 8}\left[{4-x\over 1-x}+{3x\over (1-x)^2}\ln x\right]
\end{equation}
and 
\begin{eqnarray}\label{yy1}
Y_1(x) = &&{10x + 10 x^2 + 4x^3 \over 3(1-x)^2} -
           {2x - 8x^2-x^3-x^4\over (1-x)^3} \ln x\nonumber\\
         &+&{2x - 14x^2 + x^3 - x^4\over 2(1-x)^3} \ln^2 x
           + {2x + x^3\over (1-x)^2} L_2(1-x)\nonumber\\
         &+&8x {\partial Y_0(x) \over \partial x} \ln x_\mu
\end{eqnarray}
Similarly to (\ref{xxt}) one may write
\begin{equation}\label{yyt}
Y(x_t)=\eta_Y\cdot Y_0(x_t) \qquad \eta_Y=1.012
\end{equation} 
The RG expression $Y_{NL}$ representing the charm contribution reads
\begin{equation}\label{ynl} Y_{NL}=C_{NL}-B^{(-1/2)}_{NL}  \end{equation}
where $C_{NL}$ is the $Z^0$-penguin part given in (\ref{cnln}) and
$B^{(-1/2)}_{NL}$ is the box contribution in the charm sector, relevant
for the case of final state leptons with weak isospin $T_3=-1/2$.
\begin{table}[htb]
\caption[]{The functions $Y_{\rm NL}$ and $P_0(Y)=Y_{\rm NL}/\lambda^4$
for various $\Lms^{(4)}$ and $\mc$.
\label{tab:ynlnum}}
\begin{center}
\begin{tabular}{|c|c|c|c|c|c|c|}
\hline
& \multicolumn{3}{c|}{$Y_{\rm NL}/10^{-4}$} &
  \multicolumn{3}{c|}{$P_0(Y)$} \\
\hline
$\Lms^{(4)}\ [\mev]\;\backslash\;\mc\ [\gev]$ &
1.25 & 1.30 & 1.35 & 1.25 & 1.30 & 1.35 \\
\hline
245 & 2.75  & 2.94  & 3.13 & 0.117 & 0.125 & 0.134 \\
285 & 2.80  & 2.99  & 3.19 & 0.119 & 0.128 & 0.136 \\
325 & 2.84  & 3.04  & 3.24 & 0.121 & 0.130 & 0.138 \\
365 & 2.89  & 3.09  & 3.29 & 0.123 & 0.132 & 0.140 \\
405 & 2.92  & 3.13  & 3.33 & 0.125 & 0.133 & 0.142 \\
\hline
\end{tabular}
\end{center}
\end{table}
One has 
\begin{eqnarray}\label{bmnln}
\lefteqn{B^{(-1/2)}_{NL}={x(m)\over 4}K^{24\over 25}_c\left[ 3(1-K_2)\left(
 {4\pi\over\as(\mu)}+{15212\over 1875}(1-K^{-1}_c)\right)\right.}\nonumber\\
&&-\left.\ln{\mu^2\over m^2}-
  {317\over 12}+{15212\over 625}K_2+{23081\over 7500}K K_2
  \right]
\end{eqnarray}
Note the simple relation to $B^{(1/2)}_{NL}$ in (\ref{bnln}) (for
$r=0$)
\begin{equation}\label{dbnl}
B^{(-1/2)}_{NL}-B^{(1/2)}_{NL}=\frac{3}{16} x(m) K^{24\over 25}_c (K K_2-1)
\end{equation}

\section{Effective Hamiltonians for
$K_L\to\pi^0\nu\bar\nu$, $B\to X_{s,d}\nu\bar\nu$ and 
$B_{s,d}\to l^+l^-$}

With the above results it is easy to write down also the effective
hamiltonians for $K_L\to\pi^0\nu\bar\nu$, $B\to X_{s,d}\nu\bar\nu$
and $B_{s,d}\to l^+l^-$. Since only the top contribution is
important in these cases, we have
\begin{equation}\label{hxnu}
{\cal H}_{eff} = {G_F\over \sqrt 2} {\alpha \over
2\pi \sin^2 \Theta_W} V^\ast_{tn} V_{tn^\prime}
X (x_t) (\bar nn^\prime)_{V-A} (\bar\nu\nu)_{V-A} + h.c.   \end{equation}
for the decays $K_L\to\pi^0\nu\bar\nu$, $B\to X_s\nu\bar\nu$
and $B\to X_d\nu\bar\nu$, with $(\bar nn^\prime)=(\bar sd)$, $(\bar bs)$,
$(\bar bd)$, respectively. Similarly
\begin{equation}\label{hyll}
{\cal H}_{eff} = -{G_F\over \sqrt 2} {\alpha \over
2\pi \sin^2 \Theta_W} V^\ast_{tn} V_{tn^\prime}
Y (x_t) (\bar nn^\prime)_{V-A} (\bar ll)_{V-A} + h.c.   \end{equation}
for $B_s\to l^+l^-$ and $B_d\to l^+l^-$, with
$(\bar nn^\prime)=(\bar bs)$, $(\bar bd)$.
The functions $X$, $Y$ are given in (\ref{xx}) and (\ref{yy}).
  
\section{Phenomenology of $K^+\to\pi^+\nu\bar\nu$}

\subsection{General Aspects and Standard Model Prediction}

The branching fraction of $K^+\to\pi^+\nu\bar\nu$ can be written
as follows  
\begin{equation}\label{bkpnn}
B(K^+\to\pi^+\nu\bar\nu)=\kappa_+\cdot
\left[\left(\frac{{\rm Im}\lambda_t}{\lambda^5}X(x_t)\right)^2+
\left(\frac{{\rm Re}\lambda_c}{\lambda}P_0(X)+
\frac{{\rm Re}\lambda_t}{\lambda^5}X(x_t)\right)^2\right]
\end{equation}
\begin{equation}\label{kapp}
\kappa_+=r_{K^+}\frac{3\alpha^2 B(K^+\to\pi^0 e^+\nu)}{
 2\pi^2\sin^4\Theta_W}\lambda^8=4.11\cdot 10^{-11}
\end{equation}
Here $x_t=m^2_t/M^2_W$, $\lambda_i=V^*_{is}V_{id}$ and 
$r_{K^+}=0.901$ summarizes isospin breaking corrections in relating
$K^+\to\pi^+\nu\bar\nu$ to the well measured leading decay
$K^+\to\pi^0 e^+\nu$ \cite{MP}. 
In the standard parametrization $\lambda_c$
is real to an accuracy of better than $10^{-3}$. 
We remark that in writing $B(K^+\to\pi^+\nu\bar\nu)$ in the form
of (\ref{bkpnn}) a negligibly small term
$\sim(X^e_{NL}-X^\tau_{NL})^2$ has been omitted ($0.2\%$ effect on the
branching ratio).

A prediction for $B(K^+\to\pi^+\nu\bar\nu)$ in the Standard Model
can be obtained using information on kaon CP violation
($\varepsilon_K$), $|V_{ub}/V_{cb}|$ and $B-\bar B$ mixing
to constrain the CKM parameters ${\rm Re}\lambda_t$ and
${\rm Im}\lambda_t$ in (\ref{bkpnn}).
This standard analysis of the unitarity triangle is described in
more detail in \cite{BBL}. Here we present an updated prediction
using new input from the 1998 Vancouver conference
\cite{VANC98}.
We take
\begin{equation}\label{mtbk}
B_K=0.80\pm 0.15\qquad
\sqrt{\frac{m_{B_s}}{m_{B_d}}}
\frac{f_{B_s}\sqrt{B_{B_s}}}{f_{B_d}\sqrt{B_{B_d}}}< 1.2\qquad
P_0(X)=0.42\pm 0.06
\end{equation}
for the (scheme-invariant) kaon bag parameter $B_K$, 
the $SU(3)$ breaking among the matrix elements of
$B_s$--$\bar B_s$ and $B_d$--$\bar B_d$ mixing, and
$P_0(X)$ in (\ref{bkpnn}), respectively.
Next we use \cite{VANC98}
\begin{equation}\label{vcbub}
m_t=\bar m_t(m_t)=(166\pm 5){\rm GeV}\qquad
V_{cb}=0.040\pm 0.003\qquad
|V_{ub}/V_{cb}|=0.091\pm 0.016
\end{equation}
and the experimental results on $B_d-\bar B_d$ and
$B_s-\bar B_s$ mixing \cite{VANC98}
\begin{equation}\label{dmds}
\Delta M_d=(0.471\pm 0.016) {\rm ps}^{-1}\qquad
\Delta M_s > 12.4 {\rm ps}^{-1}
\end{equation}
Scanning all parameters within the above ranges one obtains
\begin{equation}\label{kpsm}
B(K^+\to\pi^+\nu\bar\nu)=(0.82\pm 0.32)\cdot 10^{-10}
\end{equation}
where the error is dominated by the uncertainties in the
CKM parameters.
Eq. (\ref{kpsm}) may be compared with the result from Brookhaven E787
\cite{E787}
\begin{equation}\label{kpex}
B(K^+\to\pi^+\nu\bar\nu)=(4.2^{+9.7}_{-3.5})\cdot 10^{-10}
\end{equation}
Clearly, within the large uncertainties, this result is
compatible with the Standard Model expectation.

\subsection{Upper bound on $B(K^+\to\pi^+\nu\bar\nu)$
from $\Delta M_d/\Delta M_s$}

Anticipating improved experimental results, the question arises of
how large a branching fraction could still be accomodated by the
Standard Model. In other words, how large would
$B(K^+\to\pi^+\nu\bar\nu)_{exp}$ have to be, in order to
unambiguously signal New Physics.
In this context we recall that the clean nature of 
$K^+\to\pi^+\nu\bar\nu$ implies a relation
between the branching ratio and CKM parameters with very good
theoretical accuracy. However, in order to constrain the
poorly known CKM quantities to predict (\ref{kpsm}), one introduces
theoretical uncertainties (related to $|V_{ub}/V_{cb}|$ or $B_K$)
that are not intrinsic to $K^+\to\pi^+\nu\bar\nu$ itself.
We would therefore like to investigate to what extent an upper bound
can be derived on $B(K^+\to\pi^+\nu\bar\nu)$, without relying on
$|V_{ub}/V_{cb}|$ or the constraint from kaon CP violation 
($\varepsilon_K$) involving $B_K$.
For this purpose we will make use of the ratio
$\Delta M_d/\Delta M_s$. This is motivated by the theoretically
fairly clean nature of this ratio and the improved lower bound
on $\Delta M_s$ (\ref{dmds}).

In terms of Wolfenstein parameters
$\lambda$, $A$, $\varrho$ and $\eta$ \cite{WO} one has \cite{BBL}
\begin{equation}\label{vtdts}
\left|\frac{V_{td}}{V_{ts}}\right|^2=\lambda^2
\frac{R^2_t}{1+\lambda^2(2\bar\varrho-1)}\qquad
R^2_t=(1-\bar\varrho)^2 + \bar\eta^2
\end{equation}
where \cite{BLO}
\begin{equation}\label{rebar}
\bar\varrho=\varrho\left(1-\frac{\lambda^2}{2}\right)\qquad
\bar\eta=\eta\left(1-\frac{\lambda^2}{2}\right)
\end{equation}
A measurement of $\Delta M_d/\Delta M_s$ determines
$R_t$ according to \cite{BBL}
\begin{equation}\label{rt}
R_t=\frac{r_{sd}}{\lambda}\sqrt{\frac{\Delta M_d}{\Delta M_s}}
\left(1-\frac{\lambda^2}{2}(1-2\bar\varrho)\right)\qquad
r_{sd}=\sqrt{\frac{m_{B_s}}{m_{B_d}}}
\frac{f_{B_s}\sqrt{B_{B_s}}}{f_{B_d}\sqrt{B_{B_d}}}
\end{equation}
The ratio of hadronic matrix elements $r_{sd}$ has been studied
in lattice QCD. As discussed in \cite{FS}--\cite{SHA}, 
the current status can be summarized by
\begin{equation}\label{rlat}
\frac{f_{B_s}\sqrt{B_{B_s}}}{f_{B_d}\sqrt{B_{B_d}}}=1.14\pm 0.08
\end{equation}
This result is based on the quenched approximation, but the related
uncertainties are expected to be moderate for the ratio (\ref{rlat}).
In the following we shall use
\begin{equation}\label{rsd}
r_{sd}=1.2\pm 0.2
\end{equation}
Since $r_{sd}=1$ in the $SU(3)$-flavour symmetry limit, it is
the difference $r_{sd}-1$ that is a priori unknown and has to be
determined by non-perturbative calculations. To be conservative,
we have assigned a $100\%$ error on the $SU(3)$-breaking
correction in (\ref{rsd}).

We may next rewrite (\ref{bkpnn}) using an improved
Wolfenstein parametrization \cite{BLO} as \cite{BBL}
\begin{equation}\label{bkpnw}
B(K^+\to\pi^+\nu\bar\nu)=\kappa_+ A^4 X^2(x_t)\frac{1}{\sigma}
\left[(\sigma\bar\eta)^2+(\varrho_0-\bar\varrho)^2\right]
\end{equation}
\begin{equation}\label{sigr}
\sigma=\left(\frac{1}{1-\frac{\lambda^2}{2}}\right)^2\qquad
\varrho_0=1+\frac{P_0(X)}{A^2 X(x_t)}
\end{equation}
Eq. (\ref{bkpnw}) defines an (almost circular) ellipse in the
$(\bar\varrho, \bar\eta)$ plane, centered at $(\varrho_0,0)$.
Now, for fixed $R_t$, the maximum possible branching ratio
occurs for $\bar\eta=0$ and is given by
\begin{equation}\label{bkmax}
B(K^+\to\pi^+\nu\bar\nu)_{max}=\frac{\kappa_+}{\sigma}
\left[P_0(X)+A^2 X(x_t)\frac{r_{sd}}{\lambda}
\sqrt{\frac{\Delta M_d}{\Delta M_s}}\right]^2
\end{equation}
from (\ref{vtdts}), (\ref{rt}) and (\ref{bkpnw}).
This equation provides a simple and transparent relation
for the maximal
$B(K^+\to\pi^+\nu\bar\nu)$ that would still be consistent with a given
value of $\Delta M_d/\Delta M_s$ in the Standard Model.
In particular, a lower bound on $\Delta M_s$ immediately translates
into an upper bound for $B(K^+\to\pi^+\nu\bar\nu)$.
We stress that (\ref{bkmax}) is theoretically very clean. All
necessary input is known and does not involve uncontrolled theoretical
uncertainties. At present the largest error comes from $r_{sd}$,
but a systematic improvement is possible within lattice gauge theory.

For these reasons (\ref{bkmax}) can serve as a clearcut test
of the Standard Model and has the potential to indicate the
presence of New Physics in a clean manner. Using
\begin{equation}\label{dmapr}
\sqrt{\frac{\Delta M_d}{\Delta M_s}} < 0.2\qquad
A< 0.89\qquad
P_0(X) < 0.48\qquad
X(x_t) < 1.57\qquad
r_{sd} < 1.4
\end{equation}
one finds
\begin{equation}\label{bmnum}
B(K^+\to\pi^+\nu\bar\nu)_{max}=1.67\cdot 10^{-10}
\end{equation}
This limit could be further strengthened with improved input.
(For central values of parameters the number $1.67$ in
(\ref{bmnum}) would change to $0.94$.)
However, even with present knowledge, represented by (\ref{bmnum}),
the bound is strong enough to indicate a clear conflict with the
Standard Model if $B(K^+\to\pi^+\nu\bar\nu)$ should be measured
at $2\cdot 10^{-10}$.

In table \ref{tab:rsd} we illustrate how the upper bound
on $B(K^+\to\pi^+\nu\bar\nu)$ (bound A) improves when the maximal
possible value for $r_{sd}$ is assumed to be smaller.
The outcome is compared with the result from the standard analysis
of the unitarity triangle (bound B).
We find that as long as $r_{ds}$ is higher than 1.4, the $r_{ds}$ bound 
has no impact on the maximal value in the standard analysis.
We also observe that bound A, which is theoretically very reliable,
is only slightly weaker than the less clean result of
bound B.

\begin{table}[htb]
\caption[]{$B(K^+\to\pi^+\nu\bar\nu)_{max}\cdot 10^{10}$
from $\Delta M_d/\Delta M_s$ alone (bound A), compared with
the maximum value from the standard analysis (bound B), 
where also the information from $|V_{ub}/V_{cb}|$, 
$\varepsilon_K$ and $\Delta M_d$ is used. The bounds are shown for 
various $r_{sd,max}$, the maximum of the $SU(3)$ breaking parameter
$r_{sd}$.
\label{tab:rsd}}
\begin{center}
\begin{tabular}{|c|c|c|}
\hline
$r_{sd,max}$ & Bound A & Bound B\\
\hline
\hline
1.40 & 1.67 & 1.32\\
\hline 
1.30 & 1.49 & 1.21\\
\hline 
1.25 & 1.40 & 1.17\\
\hline 
1.20 & 1.32 & 1.14\\
\hline
\end{tabular}
\end{center}
\end{table}

\subsection{$|V_{td}|$ from $B(K^+\to\pi^+\nu\bar\nu)$}

Eventually, a precise experimental determination of
$B(K^+\to\pi^+\nu\bar\nu)$, in particular if compatible with
Standard Model expectations, can be used to extract
$|V_{td}|$ directly from (\ref{bkpnn}) \cite{BB6,AJB}.
We would like to illustrate such an analysis here by
detailing the sources of uncertainty and their
impact on the final result. Our findings are summarized in
table \ref{tab:vtdnum}. We remark that the sensitivity of
$|V_{td}|$ to variations in the input is fairly linear
for the parameters $B(K^+\to\pi^+\nu\bar\nu)$, and $V_{cb}$ through
$\Lms^{(4)}$, so that the effect of other choices for the
errors can be easily infered from this table.

\begin{table}[htb]
\caption[]{Sensitivity of $|V_{td}|$, extracted from 
$K^+\to\pi^+\nu\bar\nu$, to the branching ratio, to Standard
Model parameters, and to the renormalization scales in the
top and charm sector, $\mu_t$ and $\mu_c$ respectively. The
latter measure the residual theoretical uncertainties from perturbation
theory at NLO. $m_t$ and $m_c$ are ${\overline{MS}}$ masses.
The branching ratio is assumed to have the value shown in the table
for the purpose of illustration.  Variation of the inputs within
the indicated errors yields the quoted percentage change
in $|V_{td}|$ around $|V_{td}|=1.029\cdot 10^{-2}$, corresponding
to central parameter values.
\label{tab:vtdnum}}
\begin{center}
\begin{tabular}{|c|c|c|c|c|}
\hline
\multicolumn{3}{|c||}{$B(K^+\to\pi^+\nu\bar\nu)$} & 
$\mu_t/{\rm GeV}$ & $\mu_c/{\rm GeV}$ \\
\hline
\multicolumn{3}{|c||}{$(1.0\pm 0.1)\cdot 10^{-10}$} & 
$100$--$300$ & $1$--$3$ \\
\hline 
\multicolumn{3}{|c||}{$\pm 6.8\%$} & 
$\pm 0.5\%$ & $\pm 4.5\%$ \\
\hline
\hline
$V_{cb}$ & 
$|V_{ub}/V_{cb}|$ & $m_t/{\rm GeV}$ &
$m_c/{\rm GeV}$ & $\Lms^{(4)}/{\rm GeV}$ \\
\hline
$0.040\pm 0.003$ & $0.091\pm 0.016$ &
$166\pm 5$ & $1.30\pm 0.05$ & $0.325\pm 0.080$ \\
\hline
$\pm 7.6\%$ & $\pm 0.6\%$ & $\pm 3.5\%$ &
$\pm 2.9\%$ & $\pm 2.2\%$ \\
\hline
\end{tabular}
\end{center}
\end{table}

\section{Phenomenology of $K_L\to\pi^0\nu\bar\nu$}

The rare decay mode $K_L\to\pi^0\nu\bar\nu$ is a measure of
direct CP violation \cite{LI} and therefore of particular
interest.
Using the effective hamiltonian (\ref{hxnu}) and summing over three
neutrino flavours one finds \cite{BB6,BBL}
\begin{equation}\label{bklpn}
B(K_L\to\pi^0\nu\bar\nu)=\kappa_L\cdot
\left(\frac{{\rm Im}\lambda_t}{\lambda^5}X(x_t)\right)^2
\end{equation}
\begin{equation}\label{kapl}
\kappa_L=\kappa_+ \frac{r_{K_L}}{r_{K^+}}
\frac{\tau(K_L)}{\tau(K^+)}=1.80\cdot 10^{-10}
\end{equation}
with $r_{K_L}=0.944$ the isospin breaking correction from \cite{MP} and 
$\kappa_+$ given in (\ref{kapp}). Using the improved Wolfenstein
parametrization \cite{BLO} we can rewrite (\ref{bklpn}) as \cite{BBL}
\begin{equation}\label{bklpnwol1}
B(K_L\to\pi^0\nu\bar\nu)=\kappa_L \eta^2 A^4 X^2(x_t)
\end{equation}
Using the same procedure (standard analysis) and input as for 
the derivation of (\ref{kpsm}), one obtains
\begin{equation}\label{klsm}
B(K_L\to\pi^0\nu\bar\nu)=(3.1\pm 1.3)\cdot 10^{-11}
\end{equation}
Again the error comes almost entirely from the uncertainties
in the CKM parameters.

As in the case of $K^+\to\pi^+\nu\bar\nu$ discussed above,
one may ask the question about the maximal
$K_L\to\pi^0\nu\bar\nu$ branching ratio that could be tolerated
within the Standard Model. A value of $4.4\cdot 10^{-11}$ is
suggested by (\ref{klsm}). Relying on the standard analysis of
the CKM matrix, this result necessarily involves hadronic
uncertainties (related for instance to 
$\varepsilon_K$ and $|V_{ub}/V_{cb}|$).
It would be desirable to have a relation that could yield comparable
information, but based only on quantities with very good
theoretical control.
A promising possibility is again $\Delta M_d/\Delta M_s$,
in conjunction with $\sin 2\beta$, to be obtained from the
CP asymmetry in $B_d(\bar B_d)\to J/\Psi K_S$. For all practical
purposes the latter quantity has no theoretical uncertainty, like
the $K_L\to\pi^0\nu\bar\nu$ branching fraction itself.
Using the CKM relation \cite{AJB}
\begin{equation}\label{ertsb}
\bar\eta^2=\frac{R^2_t}{2}\left(1-\sqrt{1-\sin^2 2\beta}\right)
\end{equation}
together with (\ref{rt}) gives
\begin{equation}\label{bklms}
B(K_L\to\pi^0\nu\bar\nu)=\sigma\kappa_L A^4 X^2(x_t)
\frac{r^2_{sd}}{2\lambda^2}\frac{\Delta M_d}{\Delta M_s}
\left(1-\sqrt{1-\sin^2 2\beta}\right)
\end{equation}
Here we have dropped the small ${\cal O}(\lambda^2)$ term in
(\ref{rt}) for simplicity. It could always be included if a still
higher accuracy should be required. In any case, the omission of
this correction is conservative when we are interested in an
upper limit on $B(K_L\to\pi^0\nu\bar\nu)$ 
(unless $\bar\varrho > 0.5$).
A useful numerical representation of (\ref{bklms}) is
\begin{eqnarray}\label{bklnum}
B(K_L\to\pi^0\nu\bar\nu) &=& 12.0\cdot 10^{-11}
\left(\frac{V_{cb}}{0.04}\right)^4
\left(\frac{m_t}{166\,{\rm GeV}}\right)^{2.3}
\left(\frac{r_{sd}}{1.2}\right)^2
\frac{\Delta M_d/\Delta M_s}{0.04} \nonumber \\
 && \cdot\left(1-\sqrt{1-\sin^2 2\beta}\right)
\end{eqnarray}
Experimental results for $\Delta M_d/\Delta M_s$ and
$\sin 2\beta$ immediately translate into a prediction for
$B(K_L\to\pi^0\nu\bar\nu)$. If only upper bounds are available
for $\Delta M_d/\Delta M_s$ and $\sin 2\beta$, still a clean upper bound
on $B(K_L\to\pi^0\nu\bar\nu)$ can be obtained.

At present, with $\Delta M_d/\Delta M_s < 0.04$ and no information
on CP violation in B decays ($\sin 2\beta\leq 1$), this bound is
not yet very strong as can be seen from (\ref{bklnum}).
Note however that the last factor in (\ref{bklms}), (\ref{bklnum})
decreases rapidly when $\sin 2\beta$ is restricted to values
smaller than unity. Improved results on $\Delta M_d/\Delta M_s$
will also contribute to make this upper bound competitive with
the standard analysis (\ref{klsm}).
For example, taking $V_{cb}< 0.043$, $m_t < 171\,{\rm GeV}$,
$r_{sd} < 1.4$, $\Delta M_d/\Delta M_s < 0.04$ and assuming
$\sin 2\beta < 0.7$ yields
\begin{equation}\label{bklb}
B(K_L\to\pi^0\nu\bar\nu) < 6.7 \cdot 10^{-11}
\end{equation}
Let us emphasize again that the main virtues of (\ref{bklms})
with respect to the standard CKM analysis are the very high
reliability of the theoretical input and, in addition, the simple and
transparent form of this relation. Both features will be of great
advantage in trying to pin down any possible inconsistencies of the
Standard Model. A window for New Physics will exist above
the largest possible Standard Model value implied by (\ref{bklms}) 
and the model-independent upper bound \cite{GN}
\begin{equation}\label{gnb}
B(K_L\to\pi^0\nu\bar\nu) < 4.4\cdot B(K^+\to\pi^+\nu\bar\nu) < 
6.1\cdot 10^{-9}
\end{equation}
Recent discussions on specific scenarios that might lead to
an enhancement of $K\to\pi\nu\bar\nu$ branching ratios over
their Standard Model expectations can be found in
\cite{BUR} -- \cite{CI}.
As pointed out recently in \cite{BS}, the CP-violating
ratio $\varepsilon'/\varepsilon$ plays in spite of large
theoretical uncertainties a significant role in bounding possible
enhancements.

Next, we would like to point out an independent upper
limit on $B(K_L\to\pi^0\nu\bar\nu)$, which follows directly
from the upper limit on $|V_{ub}|$, using
$\eta\leq |V_{ub}/V_{cb}|/\lambda$.
It reads ($\kappa_L/\lambda^{10}=6.78\cdot 10^{-4}$)
\begin{equation}\label{klvub}
B(K_L\to\pi^0\nu\bar\nu) < \frac{\kappa_L}{\lambda^{10}}
\left| V_{ub} V_{cb} X(x_t)\right|^2 
< 6.5\cdot 10^{-11}
\end{equation}
Although still affected by the theoretical uncertainties in extracting
$|V_{ub}|$, this very clear and direct bound will become
increasingly useful as our understanding of $|V_{ub}|$ improves.

Finally, we remark that once $K_L\to\pi^0\nu\bar\nu$ and
$K^+\to\pi^+\nu\bar\nu$ have been measured, the direct extraction
of $\sin 2\beta$ and ${\rm Im}\lambda_t$ from these modes becomes
possible and offers additional opportunities for accurate tests
of flavour physics \cite{BB6}.

\section{Conclusions}

The Standard Model predicts a characteristic pattern of
rare, flavour-changing neutral current processes.
Different transitions are interrelated in particular through
the underlying common CKM structure of quark mixing.
To make decisive tests of the flavour sector it is mandatory
to have firm control over all theoretical uncertainties.

In the present paper we have discussed specific examples of
relations that rely only on theoretically clean observables.  
For instance, knowledge of $\Delta M_d/\Delta M_s$ and $\sin 2\beta$
determines $B(K_L\to\pi^0\nu\bar\nu)$. An upper limit on
$\Delta M_d/\Delta M_s$ alone yields a stringent upper limit
on $B(K^+\to\pi^+\nu\bar\nu)$.
These constraints are both simpler and theoretically cleaner than
the results of a standard analysis, where CKM parameters are
determined using $\varepsilon_K$, $|V_{ub}/V_{cb}|$ and
$\Delta M_d$. Such an analysis is at present still very useful to
obtain benchmark predictions for rare decays. We have updated
previous analyses for $K\to\pi\nu\bar\nu$ \cite{AJB}, obtaining
\begin{eqnarray}
B(K^+\to\pi^+\nu\bar\nu) &=& (0.82\pm 0.32)\cdot 10^{-10} \nonumber \\
B(K_L\to\pi^0\nu\bar\nu) &=& (3.1\pm 1.3)\cdot 10^{-11} \nonumber
\end{eqnarray}
However, in the future one might want to avoid the hadronic
uncertainties necessarily affecting these estimates, using
instead the relations mentioned above.
We note especially that a very clean, and conservative, upper
bound 
\begin{eqnarray}
B(K^+\to\pi^+\nu\bar\nu) &<& 1.67\cdot 10^{-10} \nonumber 
\end{eqnarray}
can be deduced with present knowledge from 
$\Delta M_d/\Delta M_s$ alone. Higher values would clearly be
in conflict with the Standard Model. It will be interesting to
follow the future development of the experimental result
on $B(K^+\to\pi^+\nu\bar\nu)$ in (\ref{kpex}).

The clean constraints on $B(K^+\to\pi^+\nu\bar\nu)$
and $B(K_L\to\pi^0\nu\bar\nu)$ we have discussed should be
useful to guide the interpretation of future experimental data
and might well be essential to establish, with confidence, the
existence of New Physics.

\section*{Acknowledgements}

We thank Miko\l aj Misiak and J{\"o}rg Urban for pointing out the missing
contribution to the top-quark box diagram (fig. 2c) of
\cite{BB2}, for communicating 
to us the results of ref. \cite{MU} prior to publication,
useful discussions and a careful reading of the present manuscript.

\section*{Appendix}
\newcounter{zahler}
\renewcommand{\theequation}{\Alph{zahler}.\arabic{equation}}
\setcounter{zahler}{1}
\setcounter{equation}{0}

\subsection*{Top Contribution}

In the following we discuss the origin of the additional
term that needs to be included in our previous results
in \cite{BB2}.
It arises from the box-diagram contribution to the
effective hamiltonian and is related to the existence of an
evanescent operator (non-vanishing only in $D\not= 4$ dimensions)
within the framework of dimensional regularization,
as pointed out in $\cite{MU}$. In the following argument we will
use a gluon mass to regulate infrared (IR) divergences. A different
IR regulator (external quark masses) was used in \cite{MU}. Both
approaches lead to the same result. 

To be specific we first concentrate on the top contribution
for the rare decays with neutrinos in the final state.
If we define
\begin{equation}\label{qbar}
\bar Q=\bar s\gamma^\mu\gamma^\alpha\gamma^\nu(1-\gamma_5)d
       \bar\nu\gamma_\mu\gamma_\alpha\gamma_\nu(1-\gamma_5)\nu
\end{equation}
\begin{equation}\label{qq}
Q=\bar s\gamma^\mu(1-\gamma_5)d \bar\nu\gamma_\mu(1-\gamma_5)\nu
\end{equation}
we may write
\begin{equation}\label{qbqe}
\bar Q\equiv (16+a\varepsilon)Q+E
\end{equation}
which defines the evanescent operator $E$. Here
$\varepsilon=(4-D)/2$ and $a$ is an arbitrary constant number.
To lowest order in $\alpha_s$ the one-loop box graph is simply
proportional to $\bar Q$. It therefore contributes to the
hamiltonian (in $D$ dimensions)
\begin{equation}\label{hbox}
{\cal H}_{box}\sim\frac{1}{16}B_0 \bar Q=
  B Q+B_E E
\end{equation}
where
\begin{equation}\label{bb0e}
B=B_0+{\cal O}(\alpha_s, \varepsilon)
\qquad B_E=\frac{1}{16}B_0+{\cal O}(\alpha_s)
\end{equation}
Writing $\vec B^T=(B, B_E)$, $\vec Q^T=(Q, E)$, and including
ultraviolet (UV) coun\-ter\-terms, (\ref{hbox}) becomes
\begin{equation}\label{hbct}
{\cal H}_{box}\sim \vec B^T\vec Q+\vec B^T(Z_2 Z^{-1}-1)\vec Q
\end{equation}
Here $Z_2$ is the quark-field renormalization constant and $Z$
the renormalization constant matrix of $\vec Q$ defined by
\begin{equation}\label{q0zq}
\langle\vec Q\rangle^{(0)}=Z^{-1}_2 Z\langle\vec Q\rangle
\end{equation}
relating unrenormalized ($^{(0)}$) and renormalized amputated
Green functions with operator $\vec Q$ insertion.

To ${\cal O}(\alpha_s)$ the matrix $Z$ can be calculated by considering
one-gluon exchange across the quark current in $Q$ and $E$.
To extract the UV renormalization, we may for instance treat the
external quarks as massless and on-shell and use a gluon mass 
$\lambda$ as infrared regulator. In this case
\begin{equation}\label{z2}
Z_2=1-\frac{\alpha_s}{4\pi}\frac{4}{3}\left(\frac{1}{\varepsilon}
-\gamma+\ln 4\pi+\ln\frac{\mu^2}{\lambda^2}-\frac{1}{2}\right)
\end{equation}
and the matrix $Z$ is found to be
\begin{equation}\label{zrs}
Z=1+\frac{\alpha_s}{4\pi}
\left(\begin{array}{cc}
0 & 0\\
r & s
\end{array}\right)
\qquad r=32 
\end{equation}
$Z$ implies a finite renormalization of $(Q, E)$.
Since the quark component in $Q$ is a conserved current, $Q$ by
itself is not renormalized in (\ref{zrs}). On the other hand,
$E$ mixes back into $Q$ under finite renormalization $\sim r$.
Note that $r$ is a pure short-distance quantity. It arises when the
ultraviolet pole of the loop integration $\sim 1/\varepsilon$
multiplies a term $\sim\varepsilon Q$ from the $D$-dimensional
Dirac algebra. Note also that the arbitrary number $a$ in
(\ref{qbqe}) does not appear in relation (\ref{zrs}) and is unimportant
in the following. The element $s$ in (\ref{zrs}) is likewise irrelevant
in the present context.
The hamiltonian in (\ref{hbct}) may then be rewritten as
\begin{equation}\label{hbr}
{\cal H}_{box}\sim \left(B-\frac{\alpha_s}{4\pi}r B_E\right)Z_2 Q
+B_E Z_2 E+\ldots
\end{equation}
where irrelevant terms have been omitted.

To extract the Wilson coefficient $B$, the matrix element of the
effective theory (\ref{hbr}) has to be matched to the corresponding
(renormalized) amplitude in the full theory, $Z_2 F$. If we denote
the tree-level matrix element of operator $O$ as
$\langle O\rangle_T$ and use a gluon mass $\lambda$ as infrared 
regulator, we have
\begin{equation}
Z_2\langle Q\rangle^{(0)}=\langle Q\rangle_T\qquad
Z_2\langle E\rangle^{(0)}=\frac{\alpha_s}{4\pi}
r\langle Q\rangle_T+\ldots
\end{equation}
and
\begin{equation}\label{flb}
Z_2 F^{(\lambda)}\langle Q\rangle_T=B\langle Q\rangle_T
\end{equation}
In this case the contribution of $r$ has canceled on the r.h.s.
of (\ref{hbr}) and the coefficient $B$ is given directly from
the calculation in the full theory, $B=Z_2 F^{(\lambda)}$.
We see that the presence of the evanescent operator is then
irrelevant and can be ignored from the beginning. This is not
unexpected, because the l.h.s. of (\ref{flb}) is finite after 
renormalization and the limit $D\to 4$ can be taken without the
need to specify the subtraction of evanescent operators
(in contrast to the case of the ordinary non-leptonic hamiltonian
\cite{BW}). This situation was implicitly assumed in the
calculation of the ${\cal O}(\alpha_s)$ box contribution
to the Wilson coefficient in \cite{BB2}.
However, in \cite{BB2} IR divergences were regulated dimensionally,
rather than by a gluon mass. This is a legitimate procedure, but
it involves a subtlety that was not taken into account in the
original work \cite{BB2}. 

In the case of dimensional regularization of both UV and IR
divergences, and with external quarks massless and on-shell,
$Z_2=1$ and the ${\cal O}(\alpha_s)$ corrections to the matrix
elements of the operators in (\ref{hbr}) vanish identically.
The matching relation between full and effective theory (\ref{hbr})
becomes, instead of (\ref{flb})
\begin{equation}\label{feb}
F^{(\varepsilon)}\langle Q\rangle_T=
\left(B-\frac{\alpha_s}{4\pi}r B_E\right)\langle Q\rangle_T
\end{equation}
The relevant coefficient $B$ is therefore not given by the
full theory amplitude alone (unlike the case of (\ref{flb})), but
by $F^{(\varepsilon)}+\alpha_s/(4\pi) r B_E$.
With the QCD corrected box function written as \cite{BB2}
\begin{equation}\label{bb1}
B(x,1/2)=B_0(x)+\frac{\alpha_s}{4\pi}B_1(x,1/2)
\end{equation}
in the case of external neutrinos ($T_3=1/2$), this implies that an 
additional term $\Delta B_1(1/2)=r B_E=2 B_0$ has to be added to
the result in \cite{BB2}. Recalling $X=C-4B(1/2)$, this
corresponds to $\Delta X_1=-8 B_0$, which is already included
in (\ref{xx1}).
It can be checked that the correction term $\alpha_s/(4\pi) r B_E$
is identical to the (finite) difference between the full theory
amplitudes with gluon mass and dimensional regularization, respectively,
of IR singularities, i.e. $Z_2 F^{(\lambda)}-F^{(\varepsilon)}$.
Thus, identical results are obtained for $B$ independent of
the IR regulator, as it must be the case.

\subsection*{Charm Contribution}

A similar correction arises also in the charm sector. To
determine the corresponding change in the box function
we write the result of the renormalization group analysis as
\begin{eqnarray}\label{bnlg1}
\lefteqn{B^{(1/2)}_{NL}={x(m)\over 4}K^{24\over 25}_c\left[ 3(1-K_2)\left(
 {4\pi\over\as(\mu)}+{15212\over 1875}(1-K^{-1}_c)\right)\right.} \\
&&-\left.\ln{\mu^2\over m^2}-
  {r\ln r\over 1-r}+\frac{\gamma^{(1)}_{12}}{256}-\frac{155}{6}+
{15212\over 625}K_2+
\left(\frac{9353}{3750}-\frac{\gamma^{(1)}_{12}}{256}\right)K K_2
  \right] \nonumber
\end{eqnarray}
Here we have explicitly displayed the dependence on  
$\gamma^{(1)}_{12}$. The element $\gamma^{(1)}_{12}$ of the
two-loop anomalous dimension matrix for the box contribution
determines the NLO correction in $B_{NL}$ (apart from the anomalous
dimensions of $\alpha_s$ and of the charm quark mass, which are well
known; more details on the renormalization group analysis can be
found in \cite{BB3}).
$\gamma^{(1)}_{12}$ can be fixed by expanding $B_{NL}$ to first
order in $\alpha_s$ and comparing with the top contribution $B(x)$
expanded to first order in $x$. This yields
\begin{equation}\label{g121}
\gamma^{(1)}_{12}=\frac{128}{3}
\end{equation}
and one obtains (\ref{bnln}) from (\ref{bnlg1}).

We next show that the same result follows also directly
from an analysis of operator mixing. This discussion extends
the NLO calculations for the box contribution in the charm sector
of $K^+\to\pi^+\nu\bar\nu$, presented in \cite{BB3}, to include
the effect of the evanescent operator $E$ (\ref{qbqe}).

The relevant basis of operators is given by
$\vec O=(O, Q_c, E_c)^T$ with
\begin{equation}\label{ob} O=
 -i\int d^4x\ T\left((\bar sc)_{V-A}(\bar \nu l)_{V-A}\right)(x)\
  \left((\bar l\nu)_{V-A}(\bar cd)_{V-A}\right)(0)\ -
   \{c\to u\}    
\end{equation}
\begin{equation}\label{qcec}
Q_c=\frac{m^2}{g^2}Q\qquad E_c=\frac{m^2}{g^2} E
\end{equation}
where $m=\bar m_c(\mu)$ and $g$ is the QCD coupling.

The ingredients of the renormalization group analysis are the
initial values of the Wilson coefficients and the anomalous
dimensions governing the evolution from scale $\mu=M_W$ to
$\mu=m_c$. Details on the procedure may be found in \cite{BB3,BBL}.
Here we concentrate on the role played by the evanescent operator.

The initial conditions are not affected by the presence of $E_c$.
The coefficient of $E_c$ itself is ${\cal O}(\alpha_s)$, which gives
an irrelevant contribution at NLO (where the highest order terms kept are
${\cal O}(\alpha_s)$ times a physical operator).
However, $E_c$ contributes to the physical $(O, Q_c)$ sector of the
anomalous dimension matrix through operator mixing, affecting the
NLO quantity $\gamma^{(1)}_{12}$. We work with dimensional
regularization throughout, for both UV and IR divergences.
The renormalization in the $(Q_c, E_c)$ sector can be written as
\begin{eqnarray}\label{z2qc}
Z_2\langle Q_c\rangle^{(0)} &=& \langle Q_c\rangle=\langle Q_c\rangle_T \\
\label{z2ec}
Z_2\langle E_c\rangle^{(0)} &=& Z_{32}\langle Q_c\rangle + 
Z_{33}\langle E_c\rangle = \langle E_c\rangle_T 
\end{eqnarray}
Unrenormalized, renormalized and tree-level matrix elements are denoted
by $\langle\rangle^{(0)}$, $\langle\rangle$ and $\langle\rangle_T$,
respectively. Here and in the following the renormalization of $m$ and $g$
is, however, understood to be already carried out in the matrix elements
labeled by $\langle\rangle^{(0)}$.

From (\ref{zrs}) above we have
\begin{equation}\label{z3233}
Z_{32}=\frac{\alpha_s}{4\pi} r \qquad
Z_{33}=1+\frac{\alpha_s}{4\pi} s
\end{equation}
Next
\begin{eqnarray}\label{z2o}
Z_2\langle O\rangle^{(0)} &=& \langle O\rangle +
Z_{12}\langle Q_c\rangle_T + Z_{13}\langle E_c\rangle_T \\
&=& \langle O\rangle +
(Z_{12}+Z_{13}Z_{32})\langle Q_c\rangle + Z_{13}Z_{33}\langle E_c\rangle
\nonumber
\end{eqnarray}
where
\begin{equation}\label{z1213}
Z_{12}=\frac{16}{\varepsilon}\frac{\alpha_s}{4\pi}
+{\cal O}(\alpha^2_s)\qquad
Z_{13}=\frac{1}{\varepsilon}\frac{\alpha_s}{4\pi}
+{\cal O}(\alpha^2_s)
\end{equation}
The renormalization of $\vec O=(O, Q_c, E_c)^T$ may be summarized
as
\begin{equation}\label{z2zet}
Z_2\langle\vec O\rangle^{(0)}=(1+\zeta)\langle\vec O\rangle
\end{equation}
\begin{equation}\label{zeta}
\zeta=
\left(\begin{array}{ccc}
  0 & Z_{12}+Z_{13}Z_{32} & Z_{13} Z_{33} \\
  0 & 0 & 0 \\
  0 & Z_{32} & Z_{33}-1 \end{array}\right)
={\cal O}(\alpha_s)
\end{equation}
The anomalous dimension of $\vec O$ is given by
\begin{equation}\label{gazet}
\gamma=(1+\zeta)^{-1}\Gamma(1+\zeta)+
(1+\zeta)^{-1}\frac{d\zeta}{d\ln\mu}
\end{equation}
where $\Gamma={\rm diag}(0,\gamma_{22},\gamma_{22})$, with
$\gamma_{22}$ the anomalous dimension of $m^2/g^2$.
Using
\begin{equation}\label{dzo}
\frac{d}{d\ln\mu}Z_2\langle\vec O\rangle^{(0)}=
-\Gamma Z_2 \langle\vec O\rangle^{(0)},
\qquad
\frac{d}{d\ln\mu}\langle\vec O\rangle\equiv
-\gamma\langle\vec O\rangle
\end{equation}
(\ref{gazet}) follows from (\ref{z2zet}).
To see the impact of $E_c$ (entries with index '3') on
$\gamma_{12}$, we note that the quantities $\gamma_{22}$,
$\zeta$, $Z_{13}$ and $Z_{32}$ all start at ${\cal O}(\alpha_s)$.
We then find
\begin{eqnarray}\label{ga12}
\gamma_{12} &=& -\zeta_{12}\gamma_{22}+\frac{d\zeta_{12}}{d\ln\mu}
-\zeta_{13}\frac{d\zeta_{32}}{d\ln\mu} \\
\label{ga122}
&=& - Z_{12}\gamma_{22}+\frac{d Z_{12}}{d\ln\mu}
+\frac{d Z_{13}}{d\ln\mu} Z_{32}+{\cal O}(\alpha^3_s)
\end{eqnarray}
The contribution from the evanescent operator is
\begin{equation}\label{deg12}
\Delta_E\gamma_{12}=\frac{d Z_{13}}{d\ln\mu} Z_{32}=
-2 r\left(\frac{\alpha_s}{4\pi}\right)^2=
-64\left(\frac{\alpha_s}{4\pi}\right)^2
\end{equation}
where we have used (\ref{z3233}), (\ref{z1213}) and
$d\alpha_s/d\ln\mu=-2\varepsilon\alpha_s+{\cal O}(\alpha^2_s)$.
The first two terms in (\ref{ga122}) were computed in \cite{BB3}.
With
\begin{equation}
\gamma_{12}=\frac{\alpha_s}{4\pi}\gamma^{(0)}_{12}+
\left(\frac{\alpha_s}{4\pi}\right)^2\gamma^{(1)}_{12}
\end{equation}
we then obtain, using the result of \cite{BB3}
\begin{equation}
\gamma^{(1)}_{12}=\frac{320}{3}+\Delta_E\gamma^{(1)}_{12}=
\frac{128}{3}
\end{equation}
in agreement with (\ref{g121}).

\subsection*{Summary}

In the two preceding subsections
we have discussed the case with final state neutrinos
($T_3=1/2$, $X$-functions). Completely analogous arguments apply
to the case of external charged leptons ($T_3=-1/2$, $Y$-functions)
and lead to the results summarized in the main text. 
We finally collect, for the sake of clarity, all changes due to the 
additional contribution
from the evanescent operator. They affect only the box contribution.
We list the terms that
have to be added to our previous results in \cite{BB2,BB3,BBL}
for the corresponding Wilson coefficient functions.
The modifications in the top sector read explicitly
\begin{equation}\label{dxyt}
\Delta X(x)=-8B_0(x)\frac{\alpha_s}{4\pi}\qquad
\Delta Y(x)=+8B_0(x)\frac{\alpha_s}{4\pi}
\end{equation}
where $B_0$ is the leading order box function \cite{IL}
\begin{equation}
B_0(x)=\frac{1}{4}\left[\frac{x}{1-x}+\frac{x\ln x}{(1-x)^2}\right]
\end{equation}
In the charm sector one has
\begin{equation}\label{dxyc}
\Delta X_{NL}=-\frac{1}{4} x(m) K^{24\over 25}_c (K K_2-1)\qquad
\Delta Y_{NL}=+\frac{1}{4} x(m) K^{24\over 25}_c (K K_2-1)\qquad
\end{equation} 
in the notation defined in sections 2 and 3 of the main text.
For both $X_{NL}$ and $Y_{NL}$ the relevant entry of the
two-loop anomalous dimension matrix $\gamma^{(1)}_{12}$
changes by $\Delta\gamma^{(1)}_{12}=-64$.

\vfill\eject
 
\end{document}